\pdfoutput=1

\documentclass[11pt]{article}

\usepackage{ACL2023}

\usepackage{stfloats}

\usepackage{graphicx}
\graphicspath{ {./images/} }

\usepackage{times}
\usepackage{latexsym}

\usepackage[T1]{fontenc}

\usepackage[utf8]{inputenc}
\usepackage{multirow}
\usepackage{colortbl}
\usepackage{booktabs} 
\usepackage{bm}

\usepackage{microtype}

\usepackage{inconsolata}
\usepackage{array}
\usepackage{amsmath}
\usepackage{booktabs}
\usepackage{graphicx}
\usepackage{caption}
\usepackage[aboveskip=-6pt]{subcaption}

\title{PromptRank: Unsupervised Keyphrase Extraction Using Prompt}

\author{\quad Aobo Kong$^{1}$ \quad Shiwan Zhao$^{2}$ \quad Hao Chen$^{3}$ \quad Qicheng Li$^{1}$\thanks{~~Qicheng Li is the corresponding author.} \quad Yong Qin$^{1}$\\
\textbf{\quad Ruiqi Sun$^{3}$ \quad Xiaoyan Bai$^{3}$}\\
$^1$TMCC, CS, Nankai University \quad $^2$Independent Researcher\\
$^3$Enterprise \& Cloud Research Lab, Lenovo Research \\
\texttt{$^{1}$kongaobo9@163.com \quad $^{2}$zhaosw@gmail.com} \\
\texttt{$^{1}$\{liqicheng, qinyong\}@nankai.edu.cn}\\
\texttt{$^{3}$\{chenhao31, sunrq2, baixy8\}@lenovo.com}\\
}

\begin{document}

\maketitle

\begin{abstract}

The keyphrase extraction task refers to the automatic selection of phrases from a given document to summarize its core content. State-of-the-art (SOTA) performance has recently been achieved by embedding-based algorithms, which rank candidates according to how similar their embeddings are to document embeddings. However, such solutions either struggle with the document and candidate length discrepancies or fail to fully utilize the pre-trained language model (PLM) without further fine-tuning. To this end, in this paper, we propose a simple yet effective unsupervised approach, PromptRank, based on the PLM with an encoder-decoder architecture. Specifically, PromptRank feeds the document into the encoder and calculates the probability of generating the candidate with a designed prompt by the decoder. We extensively evaluate the proposed PromptRank on six widely used benchmarks. PromptRank outperforms the SOTA approach MDERank, improving the $F1$ score relatively by 34.18$\%$, 24.87$\%$, and 17.57$\%$ for 5, 10, and 15 returned results, respectively. This demonstrates the great potential of using prompt for unsupervised keyphrase extraction. We release our code at this \href{https://github.com/HLT-NLP/PromptRank}{url}.
\end{abstract}

\section{Introduction}
\label{sec: intro}

Keyphrase extraction aims to automatically select phrases from a given document that serve as a succinct summary of the main topics, assisting readers in quickly comprehending the key information, and facilitating numerous downstream tasks like information retrieval, text mining, summarization, etc. Existing keyphrase extraction work can be divided into two categories: supervised and unsupervised approaches. With the development of deep learning, supervised keyphrase extraction methods have achieved great success by using advanced architectures, such as LSTM \cite{alzaidy2019bi, sahrawat2020keyphrase} and Transformer \cite{santosh-etal-2020-sasake, nikzad2021phraseformer, martinc2022tnt}. However, supervised methods require large-scale labeled training data and may generalize poorly to new domains. Therefore, unsupervised keyphrase extraction methods, mainly including statistics-based \cite{10.1007/978-3-319-56608-5_37, CAMPOS2020257}, graph-based \cite{bougouin-etal-2013-topicrank, boudin-2018-unsupervised}, and embedding-based methods \cite{bennani-smires-etal-2018-simple, zhang-etal-2022-mderank}, are more popular in industry scenarios.

Recent advancements in embedding-based approaches have led to SOTA performances that can be further divided into two groups. The first group of methods, such as EmbedRank~\cite{bennani-smires-etal-2018-simple} and SIFRank~\cite{8954611}, embed the document and keyphrase candidates into a latent space, calculate the similarity between the embeddings of the document and candidates, then select the top-K most similar keyphrases. Due to the discrepancy in length between the document and its candidates, these approaches perform less than optimal and are even worse for long documents. To mitigate such an issue, the second kind of approach is proposed. By leveraging a pre-trained language model (PLM), MDERank \cite{zhang-etal-2022-mderank} replaces the candidate's embedding with that of the masked document, in which the candidate is masked from the original document. With the similar length of the masked document and the original document, their distance is measured, and the greater the distance, the more significant the masked candidate as a keyphrase. Though MDERank solves the problem of length discrepancy, it faces another challenge: PLMs are not specifically optimized for measuring such distances so contrastive fine-tuning is required to further improve the performance. This places an additional burden on training and deploying keyphrase extraction systems. Furthermore, it hinders the rapid adoption of large language models when more powerful PLMs emerge.

Inspired by the work CLIP~\cite{clip}, in this paper, we propose to expand the candidate length by putting them into a well-designed template (i.e., prompt). Then to compare the document and the corresponding prompts, we adopt the encoder-decoder architecture to map the input (i.e., the original document) and the output (i.e., the prompt) into a shared latent space. The encoder-decoder architecture has been widely adopted and has achieved great success in many fields by aligning the input and output spaces, including machine translation~\cite{NIPS2017_3f5ee243}, image captioning~\cite{pmlr-v37-xuc15}, etc. Our prompt-based unsupervised keyphrase extraction method, dubbed {\bf PromptRank}, can address the aforementioned problems of existing embedding-based approaches simultaneously: on the one hand, the increased length of the prompt can mitigate the discrepancy between the document and the candidate. On the other hand, we can directly leverage PLMs with an encoder-decoder architecture (e.g., T5~\cite{raffel2020exploring}) for measuring the similarity without any fine-tuning. Specifically, after selecting keyphrase candidates, we feed the given document into the encoder and calculate the probability of generating the candidate with a designed prompt by the decoder. The higher the probability, the more important the candidate.

To the best of our knowledge, PromptRank is the first to use prompt for unsupervised keyphrase extraction. It only requires the document itself and no more information is needed. Exhaustive experiments demonstrate the effectiveness of PromptRank on both short and long texts. We believe that our work will encourage more study in this direction.

The main contributions of this paper are summarized as follows:

\textbullet\ We propose PromptRank, a simple yet effective method for unsupervised keyphrase extraction which ranks candidates using a PLM with an encoder-decoder architecture. According to our knowledge, this method is the first to extract keyphrases using prompt without supervision.

\textbullet\ We further investigate the factors that influence the ranking performance, including the candidate position information, the prompt length, and the prompt content.

\textbullet\ PromptRank is extensively evaluated on six widely used benchmarks. The results show that PromptRank outperforms the SOTA approach MDERank by a large margin, demonstrating the great potential of using prompt for unsupervised keyphrase extraction.

\section{Related Work}

{\bf Unsupervised Keyphrase Extraction.} Mainstream unsupervised keyphrase extraction methods are divided into three categories~\cite{papagiannopoulou2020review}: statistics-based, graph-based, and embedding-based methods. Statistics-based methods~\cite{won2019automatic,campos2020yake} rank candidates by comprehensively considering their statistical characteristics such as frequency, position, capitalization, and other features that capture the context information. The graph-based method is first proposed by TextRank \cite{mihalcea-tarau-2004-textrank}, which takes candidates as vertices, constructs edges according to the co-occurrence of candidates, and determines the weight of vertices through PageRank. Subsequent works, such as SingleRank \cite{wan2008single}, TopicRank \cite{bougouin-etal-2013-topicrank}, PositionRank \cite{florescu-caragea-2017-positionrank}, and MultipartiteRank \cite{boudin-2018-unsupervised}, are improvements on TextRank. Recently, embedding-based methods have achieved SOTA performance. To name a few, EmbedRank \cite{bennani-smires-etal-2018-simple} ranks candidates by the similarity of embeddings between the document and the candidate. SIFRank \cite{8954611} follows the idea of EmbedRank and combines sentence embedding model SIF \cite{arora2017simple} and pre-trained language model ELMo \cite{peters-etal-2018-deep} to get better embedding representations. However, these algorithms perform poorly on long texts due to the length mismatch between the document and the candidate. MDERank \cite{zhang-etal-2022-mderank} solves the problem by replacing the embedding of the candidate with that of the masked document but fails to fully utilize the PLMs without fine-tuning. To address such problems, in this paper, we propose PromptRank which uses prompt learning for unsupervised keyphrase extraction. In addition to statistics-based, graph-based, and embedding-based methods, AttentionRank \cite{ding-luo-2021-attentionrank} calculates self-attention and cross-attention using a pre-trained language model to determine the importance and semantic relevance of a candidate within the document.

\begin{figure*}[t]
\centerline{\includegraphics[scale=1.0]{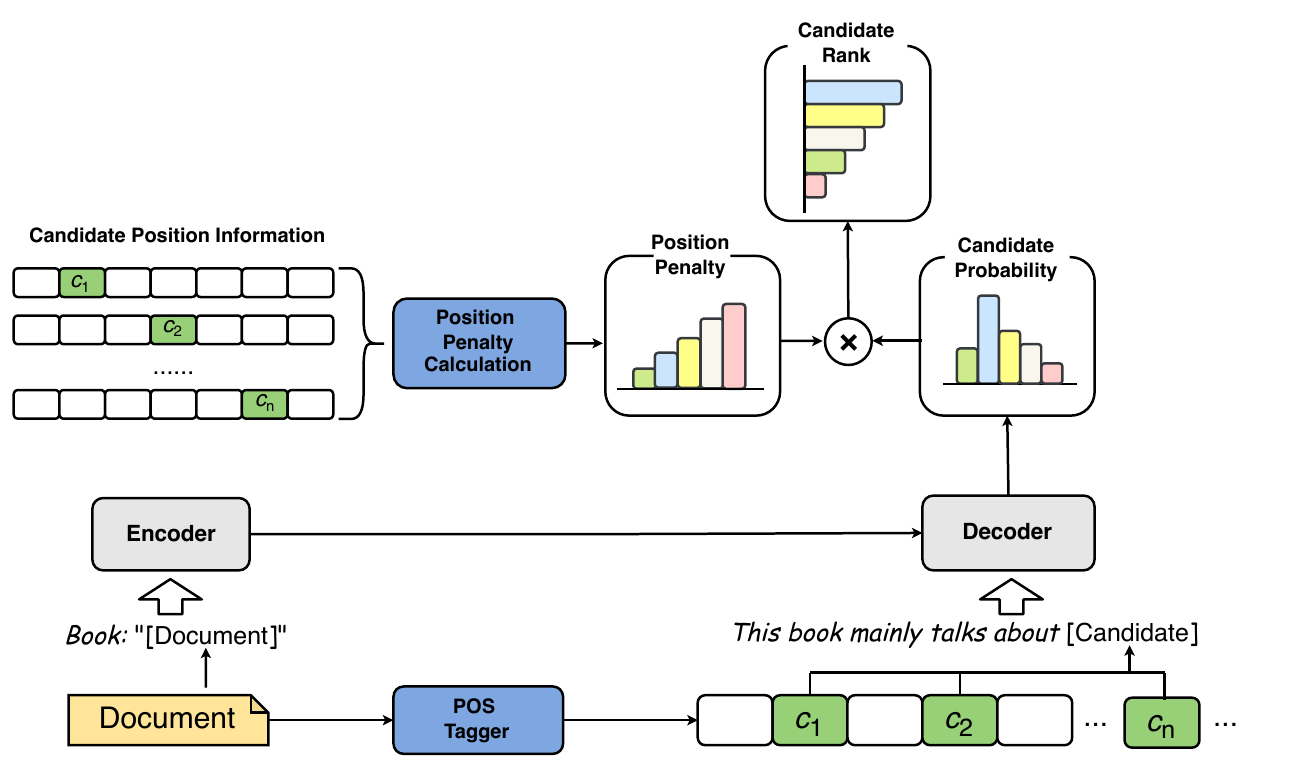}}
\caption{The core architecture of the proposed PromptRank.} 
\label{fg: architecture}
\end{figure*}

{\bf Prompt Learning.} In the field of NLP, prompt learning is considered a new paradigm to replace fine-tuning pre-trained language models on downstream tasks \cite{liu2021pre}. Compared with fine-tuning, prompt, the form of natural language, is more consistent with the pre-training task of models. Prompt-based learning has been widely used in many NLP tasks such as text classification \cite{gao-etal-2021-making, schick-schutze-2021-exploiting}, relation extraction \cite{chen2022knowprompt}, named entity recognition \cite{cui-etal-2021-template}, text generation \cite{li-liang-2021-prefix}, and so on. 
In this paper, we are the first to use prompt learning for unsupervised keyphrase extraction, leveraging the capability of PLMs with an encoder-decoder architecture, like BART \cite{lewis-etal-2020-bart} and T5 \cite{raffel2020exploring}. Our work is also inspired by CLIP~\cite{clip}, using the prompt to increase the length of candidates and alleviate the length mismatch.

\section{PromptRank}

In this section, we introduce the proposed PromptRank in detail. The core architecture of our method is shown in Figure \ref{fg: architecture}. PromptRank consists of four main steps as follows: (1) Given a document $d$, generate a candidate set $C = \{c_{1}, c_{2}, \ldots, c_{n}\}$ based on part-of-speech sequences. (2) After feeding the document into the encoder, for each candidate $c \in C$, calculate the probability of generating the candidate with a designed prompt by the decoder, denoted as $p_{c}$. (3) Use position information to calculate the position penalty of $c$, denoted as $r_{c}$. (4) Calculate the final score $s_{c}$ based on the probability and the position penalty, and then rank candidates by their $s_{c}$ in descending order.

\subsection{Candidates Generation}
\label{section: 3.1}

We follow the common practice \cite{bennani-smires-etal-2018-simple, 8954611, zhang-etal-2022-mderank} to extract noun phrases as keyphrase candidates using the regular expression <NN. *|JJ> * <NN.*> after tokenization and POS tagging.

\subsection{Probability Calculation} 

In order to address the limitations of embedding-based methods as mentioned in Section~\ref{sec: intro}, we employ an encoder-decoder architecture to transform the original document and candidate-filled templates into a shared latent space. The similarity between the document and template is determined by the probability of the decoder generating the filled template. The higher the probability, the more closely the filled template aligns with the document, and the more significant the candidate is deemed to be. To simplify the computation, we choose to place the candidate at the end of the template, so only the candidate's probability needs to be calculated to determine its rank.

A sample prompt is shown in Figure~\ref{fg: architecture}. In Section~\ref{sec: ablation}, we investigate how the length and content of the prompt affect the performance. Specifically, we fill the encoder template with the original document and fill the decoder template with one candidate at a time. Then we obtain the sequence probability $p(y_{i}\mid y_{<i})$ of the decoder template with the candidate based on PLM.
The length-normalized log-likelihood has been widely used due to its superior performance \cite{mao-etal-2019-improving, NEURIPS2020_1457c0d6, oluwatobi-mueller-2020-dlgnet}. Hence we calculate the probability for one candidate as follows: 
\begin{equation} 
\setlength{\abovedisplayskip}{3pt}
\setlength{\belowdisplayskip}{3pt}    
p_{c}=\frac{1}{({l}_{c})^{\alpha}}\sum_{i={j}}^{{j+l_{c}-1}} \log p(y_{i} \mid y_{<i}),
\label{log_p}
\end{equation} 
where $j$ is the start index of the candidate $c$, $l_{c}$ is the length of the candidate $c$, and $\alpha$ is a hyperparameter used to regulate the propensity of PromptRank towards candidate length. We use $p_{c}$ whose value is negative to evaluate the importance of candidates in descending order.

\begin{table*}
\centering
\arrayrulecolor{black}
\begin{tabular}{lcccccccccc} 
\toprule
\multirow{2}{*}{Dataset} & \multirow{2}{*}{Domain} & \multirow{2}{*}{$N_{doc}$} & \multirow{2}{*}{$L_{doc}$} & \multirow{2}{*}{$S_{can}$} & \multirow{2}{*}{$S_{gk}$} & \multicolumn{5}{c}{Gold Keyphrase Distribution}  \\ 
\cmidrule{7-11}
                         &                    &                    &                         &                    &                    & 1    & 2    & 3    & 4   & $\geq$5                     \\ 
\midrule
Inspec                   & Science            & 500                & 122                     & 15841              & 4912               & 13.5 & 52.7 & 24.9 & 6.7 & 2.2                   \\
SemEval2017              & Science            & 493                & 170                     & 21264              & 8387               & 25.7 & 34.4 & 17.5 & 8.8 & 13.6                  \\
SemEval2010              & Science            & 243                & 190                     & 4355               & 1506               & 20.5 & 53.6 & 18.9 & 4.9 & 2.1                   \\
DUC2001                  & News               & 308                & 725                     & 35926              & 2479               & 17.3 & 61.3 & 17.8 & 2.5 & 1.1                   \\
NUS                      & Science            & 211                & 7702                    & 25494              & 2453               & 26.9 & 50.6 & 15.7 & 4.6 & 2.2                   \\
Krapivin                 & Science            & 460                & 8545                    & 55875              & 2641               & 17.8 & 62.2 & 16.4 & 2.9 & 0.7                   \\
\bottomrule
\end{tabular}
\caption{Statistics of six datasets. $N_{doc}$ denotes the number of documents in each dataset. $L_{doc}$ denotes the average length of documents. $S_{can}$ and $S_{gk}$ denote the total number of candidates and gold keyphrases in each dataset, respectively. Gold Keyphrase Distribution denotes the percentage of keyphrase with different lengths in each dataset.} 
\label{tb: dataset}
\arrayrulecolor{black}
\end{table*}

\subsection{Position Penalty Calculation}
\label{sec: pos}

When writing an article, it is common practice to begin with the main points of the article.
Research has demonstrated that the position of candidates within a document can serve as an effective statistical feature for keyphrase extraction \cite{florescu-caragea-2017-positionrank, bennani-smires-etal-2018-simple, 8954611}. 

In this paper, we use a position penalty to modulate the log probability of the candidate (as shown in Equation~\ref{log_p}) by multiplication. The log probabilities are negative, so a larger value of the position penalty is assigned to unimportant positions. This results in a lower overall score for candidates in unimportant positions, reducing their likelihood of being selected as keyphrases. Specifically, for a candidate $c$, PromptRank calculates its position penalty as follows:
\begin{equation} 
r_{c}=\frac{pos}{ { len }}+\beta,
\label{eq_pos_penalty}
\end{equation} 
where $pos$ is the position of the first occurrence of $c$, $len$ is the length of the document, and $\beta$ is a parameter with a positive value to adjust the influence of position information. The larger the value of $\beta$, the smaller the role of position information in the calculation of the position penalty. That is, when $\beta$ is large, the difference in contribution to the position penalty $r_c$ between two positions will decrease. Therefore, we use different $\beta$ values to control the sensitivity of the candidate position. 

We also observe that the effectiveness of the position information correlates with the document length. The longer the article, the more effective the position information (discussed in Section~\ref{sec: ablation}). Therefore, we assign smaller value to $\beta$ for longer documents. Empirically, we formulate $\beta$ which depends on the length of the document as follows: 
\begin{equation}
\beta=\frac{\gamma}{{len}^{3}},
\label{eq_beta}
\end{equation}
where $\gamma$ is a hyperparameter that needs to be determined experimentally.

\subsection{Candidates Ranking}
After obtaining the position penalty $r_{c}$,  PromptRank calculates the final score as follows:
\begin{equation}
{ s_{c} }=r_{c} \times  { p_{c}}.
\end{equation}
The position penalty is used to adjust the log probability of the candidate, reducing the likelihood of candidates far from the beginning of the article being selected as keyphrases. We rank candidates by the final score in descending order. Finally, the top-K candidates are chosen as keyphrases.

\section{Experiments}
\label{section: 4}

\subsection{Datasets and Evaluation Metrics}
For a comprehensive and accurate evaluation, we evaluate PromptRank on six widely used datasets, in line with the current SOTA method MDERank \cite{zhang-etal-2022-mderank}. These datasets are Inspec \cite{hulth-2003-improved}, SemEval-2010 \cite{kim-etal-2010-semeval}, SemEval-2017 \cite{augenstein-etal-2017-semeval}, DUC2001 \cite{wan2008single}, NUS \cite{10.1007/978-3-540-77094-7_41}, and Krapivin \cite{krapivin2009large}, which are also used in previous works \cite{bennani-smires-etal-2018-simple, 8954611, saxena-etal-2020-keygames, ding-luo-2021-attentionrank}. The statistics of the datasets are summarized in Table~\ref{tb: dataset}. Following previous works, we use $F_{1}$ on the top 5, 10, and 15 ranked candidates to evaluate the performance of keyphrase extraction. When calculating $F{1}$, duplicate candidates will be removed, and stemming is applied.

\subsection{Baselines and Implementation Details}
We choose the same baselines as MDERank. These baselines include graph-based methods such as TextRank \cite{mihalcea-tarau-2004-textrank}, SingleRank \cite{wan2008single}, TopicRank \cite{bougouin-etal-2013-topicrank}, and MultipartiteRank \cite{boudin-2018-unsupervised}, statistics-based methods such as YAKE \cite{campos2020yake}, and embedding-based methods such as EmbedRank \cite{bennani-smires-etal-2018-simple}, SIFRank \cite{8954611}, and MDERank\cite{zhang-etal-2022-mderank} itself. We directly use the results of the baselines from MDERank. For a fair comparison, we ensure consistency in both pre-processing and post-processing of PromptRank with MDERank. We also use T5-base (220 million parameters) as our model, which has a similar scale to BERT-base \cite{devlin-etal-2019-bert} used in MDERank. Additionally, to match the settings of BERT, the maximum length for the inputs of the encoder is set to 512.

\begin{table*}[!t]
\centering

\arrayrulecolor{black}
\scalebox{0.88}{
\begin{tabular}{clccccccc} 
\toprule
\multirow{2}{*}{$F_{1}@K$}    & \multirow{2}{*}{Method} & \multicolumn{6}{c}{Dataset}                                    & \multirow{2}{*}{AVG}  \\ 
\cmidrule{3-8}
                     &                         & Inspec & SemEval2017 & SemEval2010 & DUC2001 & NUS & Krapivin   &                       \\ 
\midrule
\multirow{9}{*}{5}  & TextRank                & 21.58  & 16.43       & 7.42        & 11.02   & 1.80     & 6.04  & 10.72                 \\
                     & SingleRank              & 14.88  & 18.23       & 8.69        & 19.14   & 2.98     & 8.12  & 12.01                 \\
                     & TopicRank               & 12.20  & 17.10       & 9.93        & 19.97   & 4.54     & 8.94  & 12.11                 \\
                     & MultipartiteRank        & 13.41  & 17.39       & 10.13       & 21.70   & 6.17     & 9.29  & 13.02                 \\
                     & YAKE                    & 8.02   & 11.84       & 6.82        & 11.99   & 7.85     & 8.09  & 9.10                  \\
                     
                     & EmbedRank(BERT)         & 28.92  & 20.03       & 10.46       & 8.12    & 3.75     & 4.05  & 12.56                 \\
                     & SIFRank(ELMo)           & 29.38  & 22.38       & 11.16       & 24.30   & 3.01     & 1.62  & 15.31                 \\
                     & MDERank(BERT)           & 26.17  & 22.81       & 12.95       & 13.05   & 15.24    & 11.78 & 17.00                 \\
\cmidrule{2-9}
                     & PromptRank(T5)          & \textbf{31.73}  & \textbf{27.14}       & \textbf{17.24}       & \textbf{27.39}   & \textbf{17.24}    & \textbf{16.11} & \textbf{22.81}                 \\
                      
\midrule
\midrule
\multirow{9}{*}{10} & TextRank                & 27.53  & 25.83       & 11.27       & 17.45   & 3.02     & 9.43  & 15.76                 \\
                     & SingleRank              & 21.50  & 27.73       & 12.94       & 23.86   & 4.51    & 10.53  & 16.85                 \\
                     & TopicRank               & 17.24  & 22.62       & 12.52       & 21.73   & 7.93     & 9.01  & 15.18                 \\
                     & MultipartiteRank        & 18.18  & 23.73       & 12.91       & 24.10   & 8.57     & 9.35  & 16.14                 \\
                     & YAKE                    & 11.47  & 18.14       & 11.01       & 14.18   & 11.05     & 9.35 & 12.53                 \\
                     & EmbedRank(BERT)         & 38.55  & 31.01       & 16.35       & 11.62   & 6.34     & 6.60  & 18.41                 \\
                     & SIFRank(ELMo)           & \textbf{39.12}  & 32.60       & 16.03       & 27.60   & 5.34     & 2.52  & 20.54                 \\
                     & MDERank(BERT)           & 33.81  & 32.51       & 17.07       & 17.31   & 18.33    & 12.93 & 21.99                 \\
                               
\cmidrule{2-9}
                     & PromptRank(T5)          & 37.88  & \textbf{37.76}       & \textbf{20.66}       & \textbf{31.59}   & \textbf{20.13}    & \textbf{16.71} & \textbf{27.46 }                \\
                                
\midrule
\midrule
\multirow{9}{*}{15} & TextRank                & 27.62  & 30.50       & 13.47       & 18.84   & 3.53     & 9.95  & 17.32                 \\
                     & SingleRank              & 24.13  & 31.73       & 14.4        & 23.43   & 4.92    & 10.42  & 18.17                 \\
                     & TopicRank               & 19.33  & 24.87       & 12.26       & 20.97   & 9.37     & 8.30  & 15.85                 \\
                     & MultipartiteRank        & 20.52  & 26.87       & 13.24       & 23.62   & 10.82     & 9.16 & 17.37                 \\
                     & YAKE                    & 13.65  & 20.55       & 12.55       & 14.28   & 13.09     & 9.12 & 13.87                 \\
                     & EmbedRank(BERT)         & 39.77  & 36.72       & 19.35       & 13.58   & 8.11     & 7.84  & 20.90                 \\
                     & SIFRank(ELMo)           & \textbf{39.82}  & 37.25       & 18.42       & 27.96   & 5.86     & 3.00  & 22.05                 \\
                     & MDERank(BERT)           & 36.17  & 37.18       & 20.09       & 19.13   & 17.95    & 12.58 & 23.85                 \\
\cmidrule{2-9}
                     & PromptRank(T5)          & 38.17  & \textbf{41.57}       & \textbf{21.35}       & \textbf{31.01}   & \textbf{20.12}    & \textbf{16.02} & \textbf{28.04}                 \\
\bottomrule
\end{tabular}}
\caption{The performance of keyphrase extraction as $F_{1}@K$, $K \in \{5, 10, 15\}$ on six datasets.} 
\label{tb: results}
\arrayrulecolor{black}
\end{table*}


PromptRank is an unsupervised algorithm with only two hyperparameters to set:  $\alpha$ and $\gamma$. PromptRank is designed to have out-of-the-box generalization ability rather than fitting to a single dataset. Hence we use the same hyperparameters to evaluate PromptRank on six datasets. We set $\alpha$ to 0.6 and $\gamma$ to $1.2 \times 10^{8}$. The effects of these hyperparameters are discussed in Section~\ref{sec: ablation}.

\subsection{Overall Results}

Table~\ref{tb: results} presents the results of the $F1@5$, $F1@10$, and $F1@15$ scores for PromptRank and the baseline models on the six datasets. The results show that PromptRank achieves the best performance on almost all evaluation metrics across all six datasets, demonstrating the effectiveness of the proposed method. Specifically, PromptRank outperforms the SOTA approach MDERank, achieving an average relative improvement of 34.18$\%$, 24.87$\%$, and 17.57$\%$ for $F1@5$, $F1@10$, and $F1@15$, respectively. It is worth noting that while MDERank mainly improves the performance on two super-long datasets (Krapivin, NUS) compared to EmbedRank and SIFRank, our approach, PromptRank, achieves the best performance on almost all datasets. This highlights the generalization ability of our approach, which can work well on different datasets with different length of documents.

\begin{figure}[!t]
\centerline{\includegraphics[scale=0.45,trim=0 38 0 0,clip]{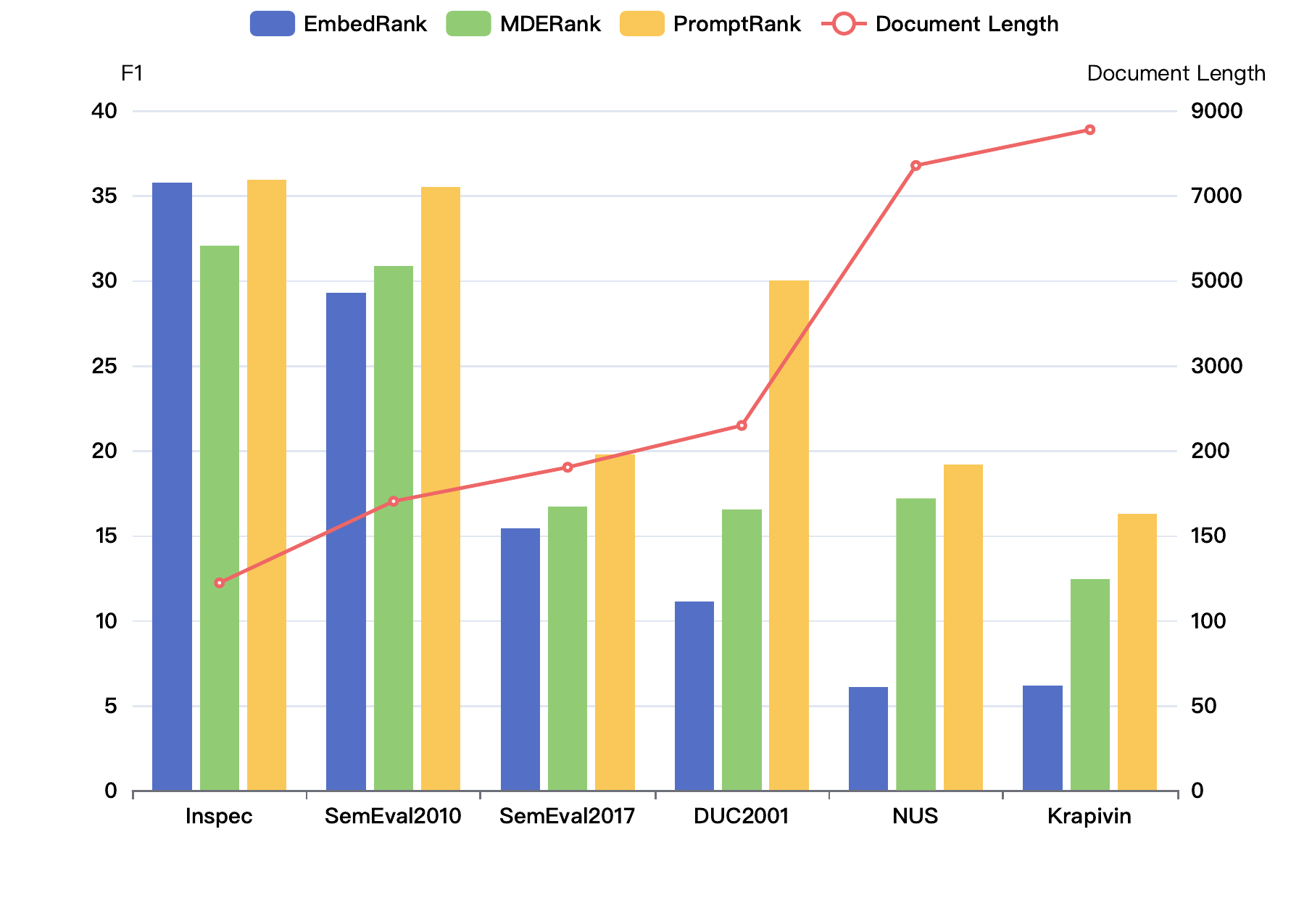}}
\caption{Performance comparison of EmbedRank, MDERank, and PromptRank as the document length increases. } 
\label{fg: comparison}
\end{figure}

As the document length increases, the length discrepancy between documents and candidates becomes more severe. To further investigate the ability of PromptRank to address this issue, we compare its performance with EmbedRank and MDERank on the average of $F1@5$, $F1@10$, $F1@15$ across the six datasets. As the length of the document increases, the number of candidates increases rapidly, and the performance of keyphrase extraction deteriorates. As shown in Figure~\ref{fg: comparison}, EmbedRank is particularly affected by the length discrepancy and its performance drops quickly. Both MDERank and PromptRank mitigate this decline. However, the masked document embedding used in MDERank does not work as well as expected. This is due to the fact that BERT is not trained to guarantee that the more important phrases are masked, the more drastically the embedding changes. BERT is just trained to restore the masked token. By leveraging a PLM of the encoder-decoder structure and using prompt, PromptRank not only more effectively solves the performance degradation of EmbedRank on long texts compared to MDERank but also performs better on short texts than both of them.


\subsection{Ablation Study}

\label{sec: ablation}

\begin{table*}
\centering
\arrayrulecolor{black}
\scalebox{0.88}{
\begin{tabular}{clccccccc} 
\bottomrule
\multirow{2}{*}{$F_{1}@K$}   & \multirow{2}{*}{Method} & \multicolumn{6}{c}{Dataset}                                     & \multicolumn{1}{c}{\multirow{2}{*}{AVG}}  \\ 
\cmidrule{3-8}
                    &                         & Inspec & SemEval2017 & SemEval2010 & DUC2001 & NUS & Krapivin   & \multicolumn{1}{l}{}                      \\ 
\midrule
\multirow{2}{*}{5}  & PromptRank$_{pt}$              & 31.79  & 27.07       & 16.74       & 23.71   & 15.81    & 14.98 & 21.68                                      \\
                    & PromptRank$_{pt+pos}$              & 31.73  & 27.14       & 17.24       & 27.39   & 17.24    & 16.11 & 22.81                                      \\
                    
\midrule
\multirow{2}{*}{10} & PromptRank$_{pt}$              & 37.84  & 37.83       & 20.82       & 28.38   & 18.99    & 16.35 & 26.70                                      \\
                    & PromptRank$_{pt+pos}$              & 37.88  & 37.76       & 20.66       & 31.59   & 20.13    & 16.71 & 27.46                                      \\
                    
\midrule
\multirow{2}{*}{15} & PromptRank$_{pt}$              & 38.17  & 41.82       & 21.15       & 28.43   & 19.59    & 15.47 & 27.44                                      \\
                    & PromptRank$_{pt+pos}$              & 38.17  & 41.57       & 21.35       & 31.01   & 20.12    & 16.02 & 28.04                                      \\
                    
\bottomrule
\end{tabular}}
\caption{The ablation study of position penalty. $pt$ represents the use of prompt-based probability. $pos$ represents the use of the position information. }
\label{tb: ablation}
\arrayrulecolor{black}
\end{table*}

{\bf Effects of Position Penalty} 
To evaluate the contribution of the position penalty to the overall performance of PromptRank, we conducted experiments in which candidates were ranked solely based on their prompt-based probability. The results are shown in Table~\ref{tb: ablation}. PromptRank without the position penalty outperforms MDERank significantly. When the position penalty is included, the performance is further improved, particularly on long-text datasets. This suggests that prompt-based probability is at the core of PromptRank, and position information can provide further benefits.

\begin{figure}[t]
\centerline{\includegraphics[scale=0.18,trim=0 0 0 0,clip]{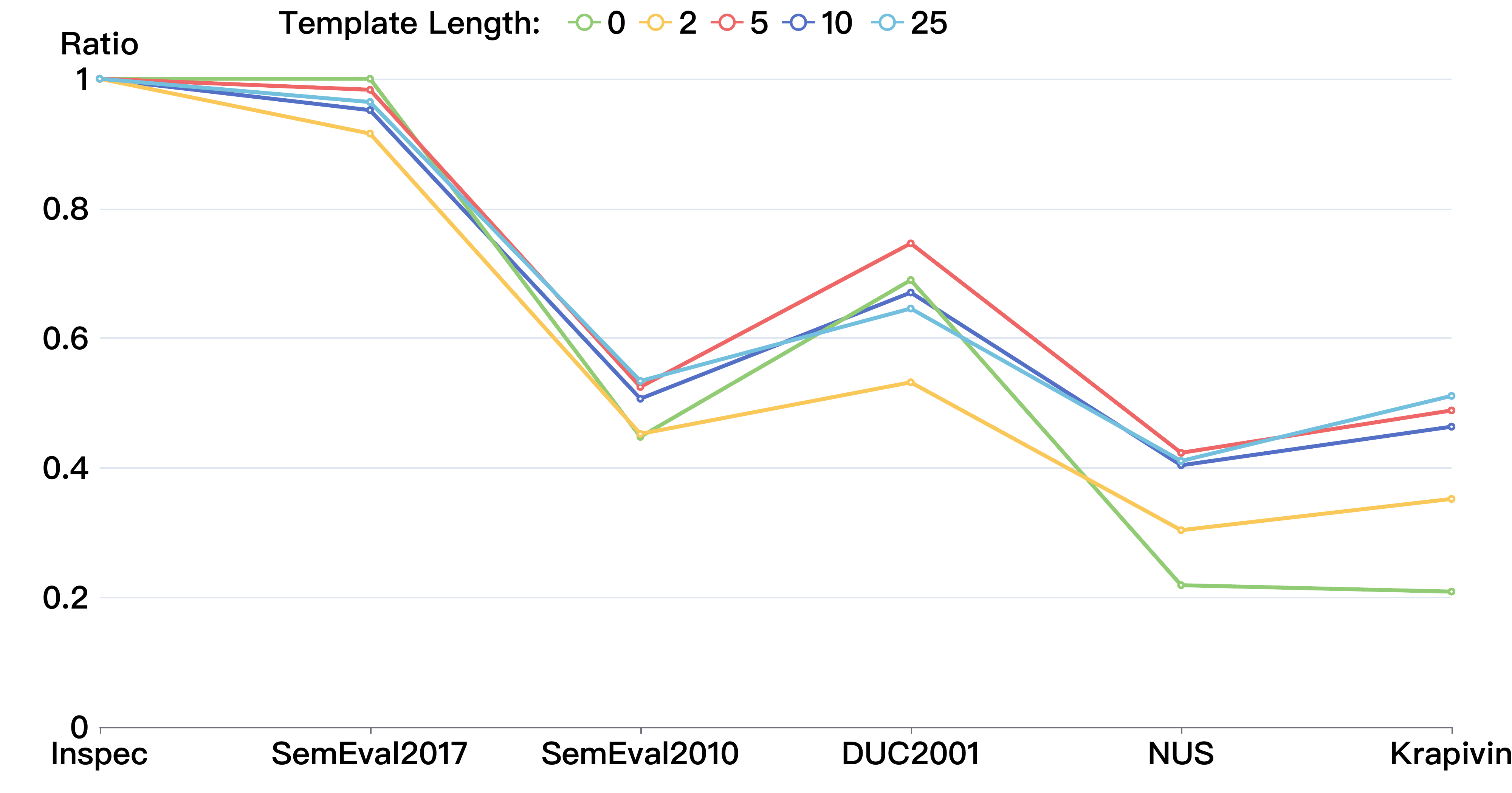}}
\caption{Comparison of the performance decay for different template lengths as the document length increases.} 
\label{fg: template-len}
\end{figure}

\noindent
{\bf Effects of Template Length} PromptRank addresses the length discrepancy of EmbedRank by filling candidates into the template. To study how long the template can avoid the drawback of EmbedRank, we conduct experiments using templates of different lengths, namely 0, 2, 5, 10, and 20. Each length contains 4 hand-crafted templates (see details in Appendix \ref{app: length}), except for the group with length 0, and the position information is not used. To exclude the impact of template content, for each template, we calculate the ratio of the performance of each dataset compared to the dataset Inspec (short text) to measure the degradation caused by an increase in text length. As shown in Figure~\ref{fg: template-len}, the higher the polyline is, the smaller the degradation is. Templates with lengths of 0 and 2 degenerate severely, facing the same problem as EmbedRank, making it difficult to exploit prompt. Templates with lengths greater than or equal to 5 better solve the length discrepancy, providing guidance for template selection.

\noindent
{\bf Effects of Template Content}
The content of the template has a direct impact on the performance of keyphrase extraction. Some typical templates and their results are shown in Table \ref{tb: prompt} (no position information used). Template 1 is empty and gets the worst results. Templates 2-5 are of the same length 5 and outperform Template 1. Template 4 achieves the best performance on all metrics. Therefore, we conclude that well-designed prompts are beneficial. Note that all templates are manually designed and we leave the automation of template construction to future work.

\begin{table*}
\centering
\arrayrulecolor{black}
\begin{tabular}{cllccc} 
\toprule
\multirow{2}{*}{Number} & \multirow{2}{*}{Encoder} & \multirow{2}{*}{Decoder}               & \multicolumn{3}{c}{$F1@K$}  \\ 
\cline{4-6}
                        &                          &                                        & 5     & 10    & 15        \\ 
\midrule
1                       & Book:"[D]"               & {[}C]                                  & 14.40 & 14.41 & 14.99     \\
2                       & Book:"[D]"               & Keywords of this book are [C]          & 14.74 & 20.02 & 21.81     \\
3                       & Book:"[D]"               & This book mainly focuses on [C] & 21.40 & 26.35 & 27.06     \\
4                       & Book:"[D]"               & This book mainly talks about [C]       &\textbf{21.69} & \textbf{26.70} & \textbf{27.44}     \\
5                       & Passage:"[D]"           & This passage mainly talks about [C]    & 21.27 & 26.15 & 27.25     \\
\bottomrule
\end{tabular}
\caption{The performance of different templates. [D] is filled with the document and [C] is filled with the candidate. F1 here is the average of six datasets.}
\label{tb: prompt}
\arrayrulecolor{black}
\end{table*}

\noindent
{\bf Effects of Hyperparameter $\bm{\alpha}$} The propensity of PromptRank for candidate length is controlled by $\alpha$. The higher $\alpha$ is, the more PromptRank tends to select long candidates. To explore the effects of different $\alpha$ values, we conduct experiments where the position information is not used. We adjust $\alpha$ from 0.2 to 1, with a step size of 0.1. The optimal values of $\alpha$ on six datasets are shown in Table \ref{tb: hyper}. $L_{gk}$ is the average number of words in gold keyphrases. Intuitively, the smaller $L_{gk}$ of the dataset, the smaller the optimal value of $\alpha$. Results show that most datasets fit this conjecture. Note that SemEval2017 with the highest $L_{gk}$ is not sensitive to $\alpha$. The reason is that the distribution of gold keyphrases in the SemEval2017 dataset is relatively more balanced (see table \ref{tb: dataset}). To maintain the generalization ability of PromptRank, it is recommended to select $\alpha$ that performs well on each benchmark rather than pursuing the best average $F1$ across all datasets. Therefore, we recommend setting the value of $\alpha$ to 0.6 for PromptRank.

\begin{table}[ht]
\centering
\arrayrulecolor{black}
\scalebox{1}{
\begin{tabular}{lcccc} 
\toprule
Dataset     & $L_{gk}$  & $\alpha$  & $\beta_{\gamma}$   \\ 
\midrule
Inspec      & 2.31 & 1         & 66.08  \\
SemEval2010 & 2.11 & 0.5       & 17.50  \\
SemEval2017 & 3.00 & 0.2$-$1         & 24.42  \\
DUC2001     & 2.07 & 0.4     & 0.89   \\
NUS         & 2.03 & 0.2     & 0.89   \\
Krapivin    & 2.07 & 0.5     & 0.89   \\
\bottomrule
\end{tabular}}
\caption{Information of hyperparameter setting. $0.2-1$ means the dataset is not sensitive to $\alpha$. $\beta_{\gamma}$ represents the average values of $beta$ calculated by $\gamma$ on various datasets and the last three datasets have the same value because of truncation. }
\label{tb: hyper}
\arrayrulecolor{black}
\end{table}

\noindent
{\bf Effects of Hyperparameter $\bm{\gamma}$} The influence of position information is controlled by $\beta$ in Equation \ref{eq_pos_penalty}. The larger the $\beta$, the smaller the impact of the position information on ranking. Previous works \cite{bennani-smires-etal-2018-simple, 8954611} show that the inclusion of position information can lead to a decrease in performance on short texts while improving performance on long texts. 
To address this, we dynamically adjust $\beta$ based on the document length through the hyperparameter $\gamma$ as shown in Equation \ref{eq_beta}, aiming to minimize the impact on short texts by a large $\beta$ while maximizing the benefits on long texts by a small $\beta$. Through experimentation, we determine the optimal value of $\gamma$ to be $1.2 \times 10^{8}$. The average values of $\beta$ calculated via $\gamma$ on six datasets are shown in Table \ref{tb: hyper}. As shown in Table \ref{tb: ablation}, the performance of PromptRank on short texts remains unchanged while performance on long texts improves significantly.

\noindent
{\bf Effects of the PLM} 
PromptRank uses T5-base as the default PLM, but to explore whether the mechanism of PromptRank is limited to a specific PLM, we conduct experiments with models of different sizes and types, such as BART \cite{lewis-etal-2020-bart}. The results, shown in Table \ref{tb: model}, indicate that even when the hyperparameters and the prompt are optimized for T5-base, the performance of all models is better than the current SOTA method MDERank. This demonstrates that PromptRank is not limited to a specific PLM and has strong versatility for different PLMs of encoder-decoder structure. 
Our approach enables rapid adoption of new PLMs when more powerful ones become available.

\begin{table}
\centering
\arrayrulecolor{black}
\begin{tabular}{lccc} 
\toprule
\multirow{2}{*}{Model} & \multicolumn{3}{c}{$F1@K$}  \\ 
\cline{2-4}
                       & 5     & 10    & 15        \\ 
\midrule
T5-small               & 21.33 & 25.93 & 26.52     \\
T5-base                & \textbf{22.81} & \textbf{27.46} & \textbf{28.04}     \\
T5-large               & 22.18 & 27.11 & 27.77     \\
BART-base              & 21.49 & 25.85 & 26.63     \\
BART-large             & 21.86 & 26.69 & 27.48     \\
\bottomrule
\end{tabular}
\caption{The performance using different PLMs. $F1$ here is the average of six datasets.}
\label{tb: model}
\arrayrulecolor{black}
\end{table}

\subsection{Case Study}

\begin{figure}[ht]
     \centering
     \begin{subfigure}[b]{0.48\textwidth}
         \centering
         \includegraphics[width=\textwidth]{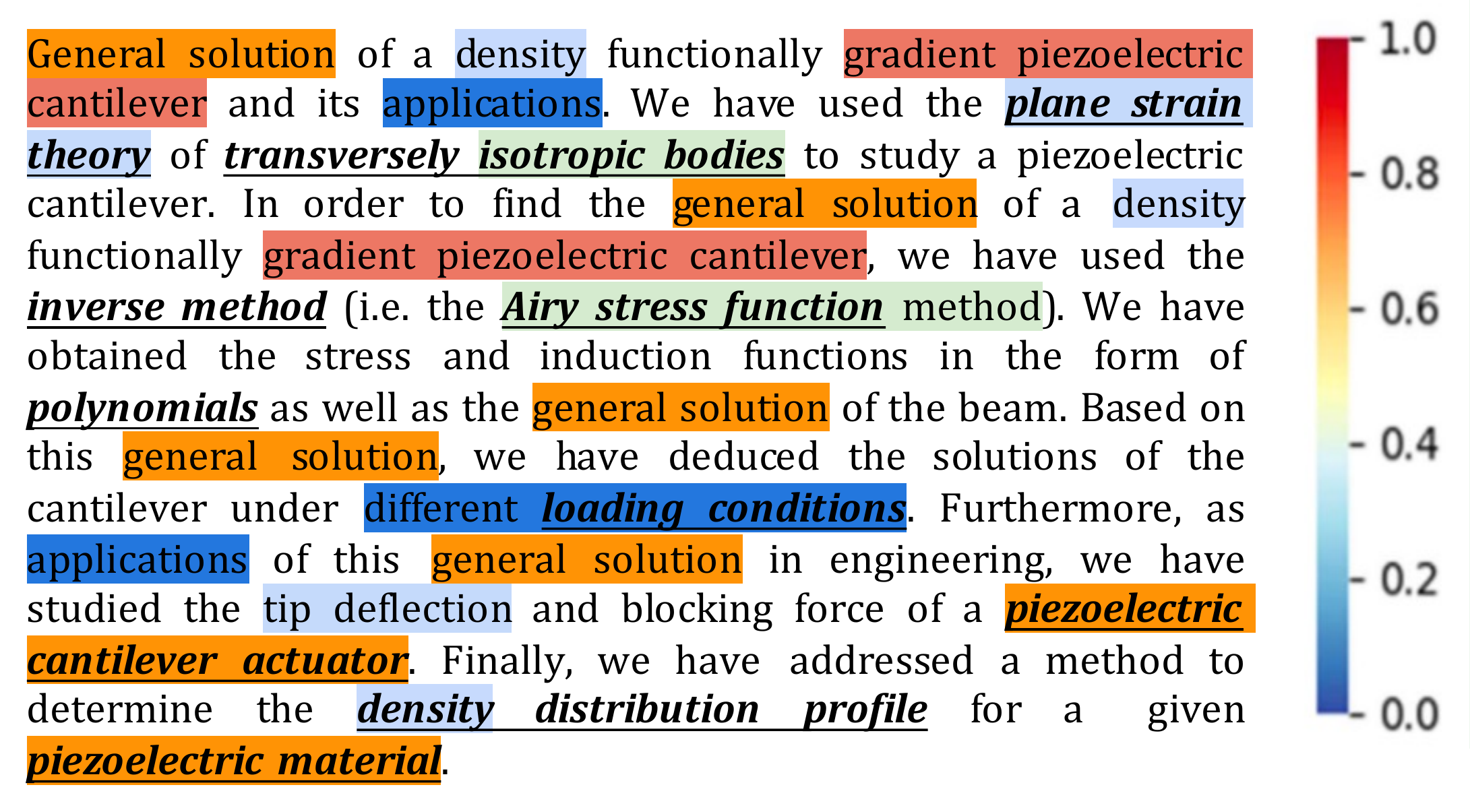}
         \caption{MDERank}
         \label{fig:y equals x}
     \end{subfigure}
    
     \begin{subfigure}[b]{0.48\textwidth}
         \centering
         \includegraphics[width=\textwidth]{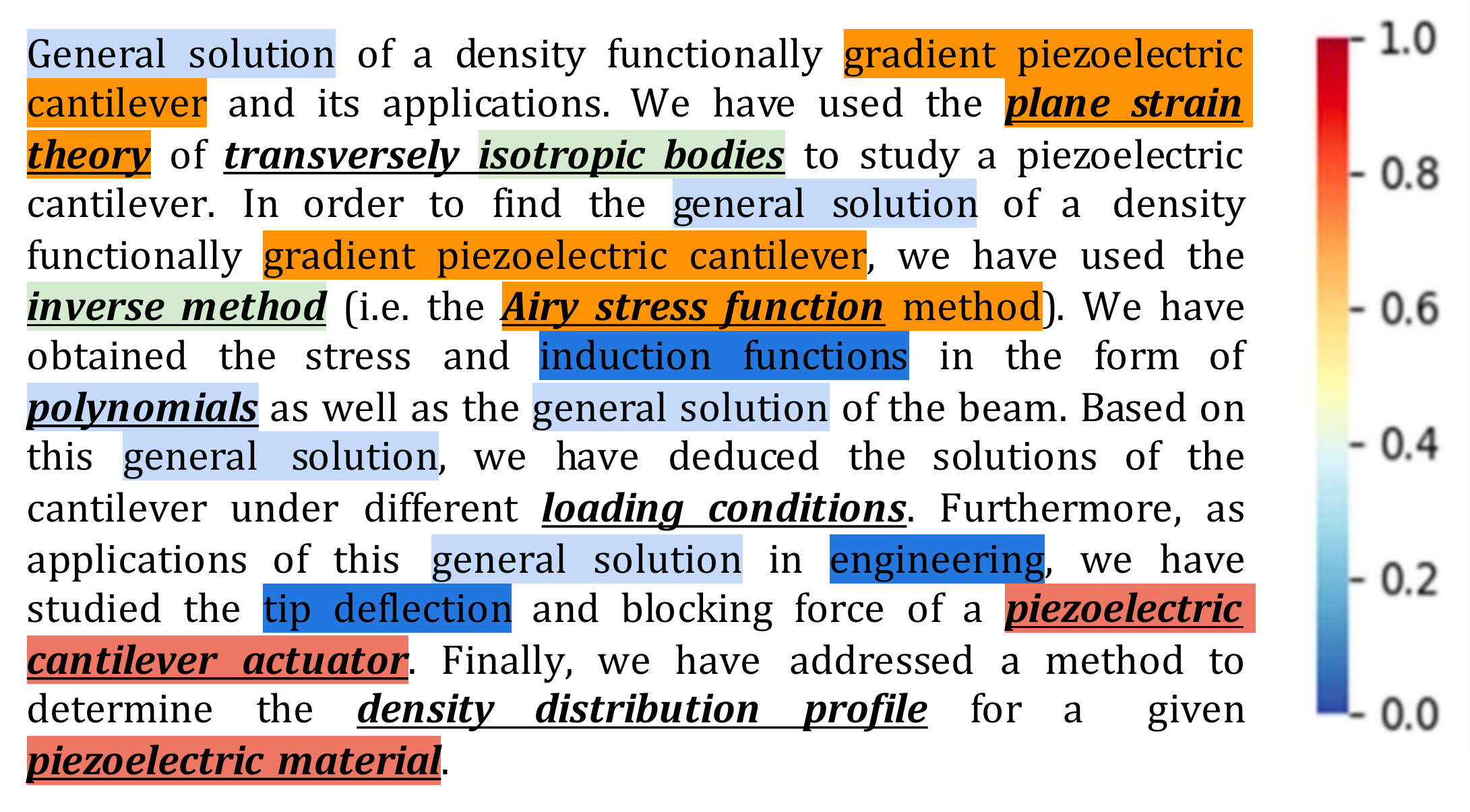}
         \caption{PromptRank}
         \label{fig:three sin x}
     \end{subfigure}
            \caption{Heat maps of candidate keyphrases by MDERank and PromptRank.}
        \label{fg: case}
\end{figure}

To demonstrate the effectiveness of PromptRank, we randomly select a document from the Inspec dataset and compare the difference between the scores produced by MDERank and PromptRank in Figure \ref{fg: case}. We normalize the original scores and present them in the form of a heat map, where the warmer the color, the higher the score, and the more important the candidate is. Gold keyphrases are underlined in bold italics. The comparison shows that compared to MDERank, PromptRank gives high scores to gold keyphrases more accurately and better distinguishes irrelevant candidates. This illustrates the improved performance of PromptRank over the SOTA method MDERank.

\section{Conclusion}
In this paper, we propose a prompt-based unsupervised keyphrase extraction method, PromptRank, using a PLM of encoder-decoder architecture. The probability of generating the candidate with a designed prompt by the decoder is calculated to rank candidates. Extensive experiments on six widely-used benchmarks demonstrate the effectiveness of our approach, which outperforms strong baselines by a significant margin. We thoroughly examine various factors that influence the performance of PromptRank and gain valuable insights. Additionally, our method does not require any modification to the architecture of PLMs and does not introduce any additional parameters, making it a simple yet powerful approach for keyphrase extraction.

\newpage

\section*{Limitations}

The core of PromptRank lies in calculating the probability of generating the candidate with a designed prompt by the decoder, which is used to rank the candidates. Our experiments have shown that the design of the prompt plays a crucial role in determining the performance of the method. While we have manually designed and selected some prompts to achieve state-of-the-art results, the process is time-consuming and may not guarantee an optimal result. To address this limitation, future research could focus on finding ways to automatically search for optimal prompts.

\section*{Acknowledgements}
The work was supported by National Key R\&D Program of China (No.2022ZD0116307), National Natural Science Foundation of China (No. 62271270) and Lenovo Research ECR lab university collaboration program.

\bibliography{anthology,custom}

\begin{thebibliography}{42}
\expandafter\ifx\csname natexlab\endcsname\relax\def\natexlab#1{#1}\fi

\bibitem[{Alzaidy et~al.(2019)Alzaidy, Caragea, and Giles}]{alzaidy2019bi}
Rabah Alzaidy, Cornelia Caragea, and C~Lee Giles. 2019.
\newblock Bi-lstm-crf sequence labeling for keyphrase extraction from scholarly
  documents.
\newblock In \emph{The world wide web conference}, pages 2551--2557.

\bibitem[{Arora et~al.(2017)Arora, Liang, and Ma}]{arora2017simple}
Sanjeev Arora, Yingyu Liang, and Tengyu Ma. 2017.
\newblock A simple but tough-to-beat baseline for sentence embeddings.
\newblock In \emph{International conference on learning representations}.

\bibitem[{Augenstein et~al.(2017)Augenstein, Das, Riedel, Vikraman, and
  McCallum}]{augenstein-etal-2017-semeval}
Isabelle Augenstein, Mrinal Das, Sebastian Riedel, Lakshmi Vikraman, and Andrew
  McCallum. 2017.
\newblock \href {https://doi.org/10.18653/v1/S17-2091} {{S}em{E}val 2017 task
  10: {S}cience{IE} - extracting keyphrases and relations from scientific
  publications}.
\newblock In \emph{Proceedings of the 11th International Workshop on Semantic
  Evaluation ({S}em{E}val-2017)}, pages 546--555, Vancouver, Canada.
  Association for Computational Linguistics.

\bibitem[{Bennani-Smires et~al.(2018)Bennani-Smires, Musat, Hossmann,
  Baeriswyl, and Jaggi}]{bennani-smires-etal-2018-simple}
Kamil Bennani-Smires, Claudiu Musat, Andreea Hossmann, Michael Baeriswyl, and
  Martin Jaggi. 2018.
\newblock \href {https://doi.org/10.18653/v1/K18-1022} {Simple unsupervised
  keyphrase extraction using sentence embeddings}.
\newblock In \emph{Proceedings of the 22nd Conference on Computational Natural
  Language Learning}, pages 221--229, Brussels, Belgium. Association for
  Computational Linguistics.

\bibitem[{Boudin(2018)}]{boudin-2018-unsupervised}
Florian Boudin. 2018.
\newblock \href {https://doi.org/10.18653/v1/N18-2105} {Unsupervised keyphrase
  extraction with multipartite graphs}.
\newblock In \emph{Proceedings of the 2018 Conference of the North {A}merican
  Chapter of the Association for Computational Linguistics: Human Language
  Technologies, Volume 2 (Short Papers)}, pages 667--672, New Orleans,
  Louisiana. Association for Computational Linguistics.

\bibitem[{Bougouin et~al.(2013)Bougouin, Boudin, and
  Daille}]{bougouin-etal-2013-topicrank}
Adrien Bougouin, Florian Boudin, and B{\'e}atrice Daille. 2013.
\newblock \href {https://aclanthology.org/I13-1062} {{T}opic{R}ank: Graph-based
  topic ranking for keyphrase extraction}.
\newblock In \emph{Proceedings of the Sixth International Joint Conference on
  Natural Language Processing}, pages 543--551, Nagoya, Japan. Asian Federation
  of Natural Language Processing.

\bibitem[{Brown et~al.(2020)Brown, Mann, Ryder, Subbiah, Kaplan, Dhariwal,
  Neelakantan, Shyam, Sastry, Askell, Agarwal, Herbert-Voss, Krueger, Henighan,
  Child, Ramesh, Ziegler, Wu, Winter, Hesse, Chen, Sigler, Litwin, Gray, Chess,
  Clark, Berner, McCandlish, Radford, Sutskever, and
  Amodei}]{NEURIPS2020_1457c0d6}
Tom Brown, Benjamin Mann, Nick Ryder, Melanie Subbiah, Jared~D Kaplan, Prafulla
  Dhariwal, Arvind Neelakantan, Pranav Shyam, Girish Sastry, Amanda Askell,
  Sandhini Agarwal, Ariel Herbert-Voss, Gretchen Krueger, Tom Henighan, Rewon
  Child, Aditya Ramesh, Daniel Ziegler, Jeffrey Wu, Clemens Winter, Chris
  Hesse, Mark Chen, Eric Sigler, Mateusz Litwin, Scott Gray, Benjamin Chess,
  Jack Clark, Christopher Berner, Sam McCandlish, Alec Radford, Ilya Sutskever,
  and Dario Amodei. 2020.
\newblock \href
  {https://proceedings.neurips.cc/paper/2020/file/1457c0d6bfcb4967418bfb8ac142f64a-Paper.pdf}
  {Language models are few-shot learners}.
\newblock In \emph{Advances in Neural Information Processing Systems},
  volume~33, pages 1877--1901. Curran Associates, Inc.

\bibitem[{Campos et~al.(2020{\natexlab{a}})Campos, Mangaravite, Pasquali,
  Jorge, Nunes, and Jatowt}]{campos2020yake}
Ricardo Campos, V{\'\i}tor Mangaravite, Arian Pasquali, Al{\'\i}pio Jorge,
  C{\'e}lia Nunes, and Adam Jatowt. 2020{\natexlab{a}}.
\newblock Yake! keyword extraction from single documents using multiple local
  features.
\newblock \emph{Information Sciences}, 509:257--289.

\bibitem[{Campos et~al.(2020{\natexlab{b}})Campos, Mangaravite, Pasquali,
  Jorge, Nunes, and Jatowt}]{CAMPOS2020257}
Ricardo Campos, Vítor Mangaravite, Arian Pasquali, Alípio Jorge, Célia
  Nunes, and Adam Jatowt. 2020{\natexlab{b}}.
\newblock \href {https://doi.org/https://doi.org/10.1016/j.ins.2019.09.013}
  {Yake! keyword extraction from single documents using multiple local
  features}.
\newblock \emph{Information Sciences}, 509:257--289.

\bibitem[{Chen et~al.(2022)Chen, Zhang, Xie, Deng, Yao, Tan, Huang, Si, and
  Chen}]{chen2022knowprompt}
Xiang Chen, Ningyu Zhang, Xin Xie, Shumin Deng, Yunzhi Yao, Chuanqi Tan, Fei
  Huang, Luo Si, and Huajun Chen. 2022.
\newblock Knowprompt: Knowledge-aware prompt-tuning with synergistic
  optimization for relation extraction.
\newblock In \emph{Proceedings of the ACM Web Conference 2022}, pages
  2778--2788.

\bibitem[{Cui et~al.(2021)Cui, Wu, Liu, Yang, and
  Zhang}]{cui-etal-2021-template}
Leyang Cui, Yu~Wu, Jian Liu, Sen Yang, and Yue Zhang. 2021.
\newblock \href {https://doi.org/10.18653/v1/2021.findings-acl.161}
  {Template-based named entity recognition using {BART}}.
\newblock In \emph{Findings of the Association for Computational Linguistics:
  ACL-IJCNLP 2021}, pages 1835--1845, Online. Association for Computational
  Linguistics.

\bibitem[{Devlin et~al.(2019)Devlin, Chang, Lee, and
  Toutanova}]{devlin-etal-2019-bert}
Jacob Devlin, Ming-Wei Chang, Kenton Lee, and Kristina Toutanova. 2019.
\newblock \href {https://doi.org/10.18653/v1/N19-1423} {{BERT}: Pre-training of
  deep bidirectional transformers for language understanding}.
\newblock In \emph{Proceedings of the 2019 Conference of the North {A}merican
  Chapter of the Association for Computational Linguistics: Human Language
  Technologies, Volume 1 (Long and Short Papers)}, pages 4171--4186,
  Minneapolis, Minnesota. Association for Computational Linguistics.

\bibitem[{Ding and Luo(2021)}]{ding-luo-2021-attentionrank}
Haoran Ding and Xiao Luo. 2021.
\newblock \href {https://doi.org/10.18653/v1/2021.emnlp-main.146}
  {{A}ttention{R}ank: Unsupervised keyphrase extraction using self and cross
  attentions}.
\newblock In \emph{Proceedings of the 2021 Conference on Empirical Methods in
  Natural Language Processing}, pages 1919--1928, Online and Punta Cana,
  Dominican Republic. Association for Computational Linguistics.

\bibitem[{Florescu and
  Caragea(2017{\natexlab{a}})}]{10.1007/978-3-319-56608-5_37}
Corina Florescu and Cornelia Caragea. 2017{\natexlab{a}}.
\newblock A new scheme for scoring phrases in unsupervised keyphrase
  extraction.
\newblock In \emph{Advances in Information Retrieval}, pages 477--483, Cham.
  Springer International Publishing.

\bibitem[{Florescu and
  Caragea(2017{\natexlab{b}})}]{florescu-caragea-2017-positionrank}
Corina Florescu and Cornelia Caragea. 2017{\natexlab{b}}.
\newblock \href {https://doi.org/10.18653/v1/P17-1102} {{P}osition{R}ank: An
  unsupervised approach to keyphrase extraction from scholarly documents}.
\newblock In \emph{Proceedings of the 55th Annual Meeting of the Association
  for Computational Linguistics (Volume 1: Long Papers)}, pages 1105--1115,
  Vancouver, Canada. Association for Computational Linguistics.

\bibitem[{Gao et~al.(2021)Gao, Fisch, and Chen}]{gao-etal-2021-making}
Tianyu Gao, Adam Fisch, and Danqi Chen. 2021.
\newblock \href {https://doi.org/10.18653/v1/2021.acl-long.295} {Making
  pre-trained language models better few-shot learners}.
\newblock In \emph{Proceedings of the 59th Annual Meeting of the Association
  for Computational Linguistics and the 11th International Joint Conference on
  Natural Language Processing (Volume 1: Long Papers)}, pages 3816--3830,
  Online. Association for Computational Linguistics.

\bibitem[{Hulth(2003)}]{hulth-2003-improved}
Anette Hulth. 2003.
\newblock \href {https://aclanthology.org/W03-1028} {Improved automatic keyword
  extraction given more linguistic knowledge}.
\newblock In \emph{Proceedings of the 2003 Conference on Empirical Methods in
  Natural Language Processing}, pages 216--223.

\bibitem[{Kim et~al.(2010)Kim, Medelyan, Kan, and
  Baldwin}]{kim-etal-2010-semeval}
Su~Nam Kim, Olena Medelyan, Min-Yen Kan, and Timothy Baldwin. 2010.
\newblock \href {https://aclanthology.org/S10-1004} {{S}em{E}val-2010 task 5 :
  Automatic keyphrase extraction from scientific articles}.
\newblock In \emph{Proceedings of the 5th International Workshop on Semantic
  Evaluation}, pages 21--26, Uppsala, Sweden. Association for Computational
  Linguistics.

\bibitem[{Krapivin et~al.(2009)Krapivin, Autaeu, and
  Marchese}]{krapivin2009large}
Mikalai Krapivin, Aliaksandr Autaeu, and Maurizio Marchese. 2009.
\newblock Large dataset for keyphrases extraction.

\bibitem[{Lewis et~al.(2020)Lewis, Liu, Goyal, Ghazvininejad, Mohamed, Levy,
  Stoyanov, and Zettlemoyer}]{lewis-etal-2020-bart}
Mike Lewis, Yinhan Liu, Naman Goyal, Marjan Ghazvininejad, Abdelrahman Mohamed,
  Omer Levy, Veselin Stoyanov, and Luke Zettlemoyer. 2020.
\newblock \href {https://doi.org/10.18653/v1/2020.acl-main.703} {{BART}:
  Denoising sequence-to-sequence pre-training for natural language generation,
  translation, and comprehension}.
\newblock In \emph{Proceedings of the 58th Annual Meeting of the Association
  for Computational Linguistics}, pages 7871--7880, Online. Association for
  Computational Linguistics.

\bibitem[{Li and Liang(2021)}]{li-liang-2021-prefix}
Xiang~Lisa Li and Percy Liang. 2021.
\newblock \href {https://doi.org/10.18653/v1/2021.acl-long.353} {Prefix-tuning:
  Optimizing continuous prompts for generation}.
\newblock In \emph{Proceedings of the 59th Annual Meeting of the Association
  for Computational Linguistics and the 11th International Joint Conference on
  Natural Language Processing (Volume 1: Long Papers)}, pages 4582--4597,
  Online. Association for Computational Linguistics.

\bibitem[{Liu et~al.(2021)Liu, Yuan, Fu, Jiang, Hayashi, and
  Neubig}]{liu2021pre}
Pengfei Liu, Weizhe Yuan, Jinlan Fu, Zhengbao Jiang, Hiroaki Hayashi, and
  Graham Neubig. 2021.
\newblock Pre-train, prompt, and predict: A systematic survey of prompting
  methods in natural language processing.
\newblock \emph{arXiv preprint arXiv:2107.13586}.

\bibitem[{Mao et~al.(2019)Mao, Majumder, McAuley, and
  Cottrell}]{mao-etal-2019-improving}
Huanru~Henry Mao, Bodhisattwa~Prasad Majumder, Julian McAuley, and Garrison
  Cottrell. 2019.
\newblock \href {https://doi.org/10.18653/v1/D19-1615} {Improving neural story
  generation by targeted common sense grounding}.
\newblock In \emph{Proceedings of the 2019 Conference on Empirical Methods in
  Natural Language Processing and the 9th International Joint Conference on
  Natural Language Processing (EMNLP-IJCNLP)}, pages 5988--5993, Hong Kong,
  China. Association for Computational Linguistics.

\bibitem[{Martinc et~al.(2022)Martinc, {\v{S}}krlj, and
  Pollak}]{martinc2022tnt}
Matej Martinc, Bla{\v{z}} {\v{S}}krlj, and Senja Pollak. 2022.
\newblock Tnt-kid: Transformer-based neural tagger for keyword identification.
\newblock \emph{Natural Language Engineering}, 28(4):409--448.

\bibitem[{Mihalcea and Tarau(2004)}]{mihalcea-tarau-2004-textrank}
Rada Mihalcea and Paul Tarau. 2004.
\newblock \href {https://aclanthology.org/W04-3252} {{T}ext{R}ank: Bringing
  order into text}.
\newblock In \emph{Proceedings of the 2004 Conference on Empirical Methods in
  Natural Language Processing}, pages 404--411, Barcelona, Spain. Association
  for Computational Linguistics.

\bibitem[{Nguyen and Kan(2007)}]{10.1007/978-3-540-77094-7_41}
Thuy~Dung Nguyen and Min-Yen Kan. 2007.
\newblock Keyphrase extraction in scientific publications.
\newblock In \emph{Asian Digital Libraries. Looking Back 10 Years and Forging
  New Frontiers}, pages 317--326, Berlin, Heidelberg. Springer Berlin
  Heidelberg.

\bibitem[{Nikzad-Khasmakhi et~al.(2021)Nikzad-Khasmakhi, Feizi-Derakhshi,
  Asgari-Chenaghlu, Balafar, Feizi-Derakhshi, Rahkar-Farshi, Ramezani,
  Jahanbakhsh-Nagadeh, Zafarani-Moattar, and
  Ranjbar-Khadivi}]{nikzad2021phraseformer}
Narjes Nikzad-Khasmakhi, Mohammad-Reza Feizi-Derakhshi, Meysam
  Asgari-Chenaghlu, Mohammad-Ali Balafar, Ali-Reza Feizi-Derakhshi, Taymaz
  Rahkar-Farshi, Majid Ramezani, Zoleikha Jahanbakhsh-Nagadeh, Elnaz
  Zafarani-Moattar, and Mehrdad Ranjbar-Khadivi. 2021.
\newblock Phraseformer: Multimodal key-phrase extraction using transformer and
  graph embedding.
\newblock \emph{arXiv preprint arXiv:2106.04939}.

\bibitem[{Oluwatobi and Mueller(2020)}]{oluwatobi-mueller-2020-dlgnet}
Olabiyi Oluwatobi and Erik Mueller. 2020.
\newblock \href {https://doi.org/10.18653/v1/2020.nlp4convai-1.7} {{DLGN}et: A
  transformer-based model for dialogue response generation}.
\newblock In \emph{Proceedings of the 2nd Workshop on Natural Language
  Processing for Conversational AI}, pages 54--62, Online. Association for
  Computational Linguistics.

\bibitem[{Papagiannopoulou and Tsoumakas(2020)}]{papagiannopoulou2020review}
Eirini Papagiannopoulou and Grigorios Tsoumakas. 2020.
\newblock A review of keyphrase extraction.
\newblock \emph{Wiley Interdisciplinary Reviews: Data Mining and Knowledge
  Discovery}, 10(2):e1339.

\bibitem[{Peters et~al.(2018)Peters, Neumann, Iyyer, Gardner, Clark, Lee, and
  Zettlemoyer}]{peters-etal-2018-deep}
Matthew~E. Peters, Mark Neumann, Mohit Iyyer, Matt Gardner, Christopher Clark,
  Kenton Lee, and Luke Zettlemoyer. 2018.
\newblock \href {https://doi.org/10.18653/v1/N18-1202} {Deep contextualized
  word representations}.
\newblock In \emph{Proceedings of the 2018 Conference of the North {A}merican
  Chapter of the Association for Computational Linguistics: Human Language
  Technologies, Volume 1 (Long Papers)}, pages 2227--2237, New Orleans,
  Louisiana. Association for Computational Linguistics.

\bibitem[{Radford et~al.(2021)Radford, Kim, Hallacy, Ramesh, Goh, Agarwal,
  Sastry, Askell, Mishkin, Clark, Krueger, and Sutskever}]{clip}
Alec Radford, Jong~Wook Kim, Chris Hallacy, Aditya Ramesh, Gabriel Goh,
  Sandhini Agarwal, Girish Sastry, Amanda Askell, Pamela Mishkin, Jack Clark,
  Gretchen Krueger, and Ilya Sutskever. 2021.
\newblock \href {http://proceedings.mlr.press/v139/radford21a.html} {Learning
  transferable visual models from natural language supervision}.
\newblock In \emph{Proceedings of the 38th International Conference on Machine
  Learning, {ICML} 2021, 18-24 July 2021, Virtual Event}, volume 139 of
  \emph{Proceedings of Machine Learning Research}, pages 8748--8763. {PMLR}.

\bibitem[{Raffel et~al.(2020)Raffel, Shazeer, Roberts, Lee, Narang, Matena,
  Zhou, Li, Liu et~al.}]{raffel2020exploring}
Colin Raffel, Noam Shazeer, Adam Roberts, Katherine Lee, Sharan Narang, Michael
  Matena, Yanqi Zhou, Wei Li, Peter~J Liu, et~al. 2020.
\newblock Exploring the limits of transfer learning with a unified text-to-text
  transformer.
\newblock \emph{J. Mach. Learn. Res.}, 21(140):1--67.

\bibitem[{Sahrawat et~al.(2020)Sahrawat, Mahata, Zhang, Kulkarni, Sharma,
  Gosangi, Stent, Kumar, Shah, and Zimmermann}]{sahrawat2020keyphrase}
Dhruva Sahrawat, Debanjan Mahata, Haimin Zhang, Mayank Kulkarni, Agniv Sharma,
  Rakesh Gosangi, Amanda Stent, Yaman Kumar, Rajiv~Ratn Shah, and Roger
  Zimmermann. 2020.
\newblock Keyphrase extraction as sequence labeling using contextualized
  embeddings.
\newblock In \emph{European Conference on Information Retrieval}, pages
  328--335. Springer.

\bibitem[{Santosh et~al.(2020)Santosh, Kumar~Sanyal, Bhowmick, and
  Das}]{santosh-etal-2020-sasake}
T.y.s.s Santosh, Debarshi Kumar~Sanyal, Plaban~Kumar Bhowmick, and
  Partha~Pratim Das. 2020.
\newblock \href {https://doi.org/10.18653/v1/2020.coling-main.469} {{S}a{SAKE}:
  Syntax and semantics aware keyphrase extraction from research papers}.
\newblock In \emph{Proceedings of the 28th International Conference on
  Computational Linguistics}, pages 5372--5383, Barcelona, Spain (Online).
  International Committee on Computational Linguistics.

\bibitem[{Saxena et~al.(2020)Saxena, Mangal, and
  Jain}]{saxena-etal-2020-keygames}
Arnav Saxena, Mudit Mangal, and Goonjan Jain. 2020.
\newblock \href {https://doi.org/10.18653/v1/2020.coling-main.184}
  {{K}ey{G}ames: A game theoretic approach to automatic keyphrase extraction}.
\newblock In \emph{Proceedings of the 28th International Conference on
  Computational Linguistics}, pages 2037--2048, Barcelona, Spain (Online).
  International Committee on Computational Linguistics.

\bibitem[{Schick and Sch{\"u}tze(2021)}]{schick-schutze-2021-exploiting}
Timo Schick and Hinrich Sch{\"u}tze. 2021.
\newblock \href {https://doi.org/10.18653/v1/2021.eacl-main.20} {Exploiting
  cloze-questions for few-shot text classification and natural language
  inference}.
\newblock In \emph{Proceedings of the 16th Conference of the European Chapter
  of the Association for Computational Linguistics: Main Volume}, pages
  255--269, Online. Association for Computational Linguistics.

\bibitem[{Sun et~al.(2020)Sun, Qiu, Zheng, Wang, and Zhang}]{8954611}
Yi~Sun, Hangping Qiu, Yu~Zheng, Zhongwei Wang, and Chaoran Zhang. 2020.
\newblock \href {https://doi.org/10.1109/ACCESS.2020.2965087} {Sifrank: A new
  baseline for unsupervised keyphrase extraction based on pre-trained language
  model}.
\newblock \emph{IEEE Access}, 8:10896--10906.

\bibitem[{Vaswani et~al.(2017)Vaswani, Shazeer, Parmar, Uszkoreit, Jones,
  Gomez, Kaiser, and Polosukhin}]{NIPS2017_3f5ee243}
Ashish Vaswani, Noam Shazeer, Niki Parmar, Jakob Uszkoreit, Llion Jones,
  Aidan~N Gomez, \L~ukasz Kaiser, and Illia Polosukhin. 2017.
\newblock \href
  {https://proceedings.neurips.cc/paper/2017/file/3f5ee243547dee91fbd053c1c4a845aa-Paper.pdf}
  {Attention is all you need}.
\newblock In \emph{Advances in Neural Information Processing Systems},
  volume~30. Curran Associates, Inc.

\bibitem[{Wan and Xiao(2008)}]{wan2008single}
Xiaojun Wan and Jianguo Xiao. 2008.
\newblock Single document keyphrase extraction using neighborhood knowledge.
\newblock In \emph{AAAI}, volume~8, pages 855--860.

\bibitem[{Won et~al.(2019)Won, Martins, and Raimundo}]{won2019automatic}
Miguel Won, Bruno Martins, and Filipa Raimundo. 2019.
\newblock Automatic extraction of relevant keyphrases for the study of issue
  competition.
\newblock In \emph{Proceedings of the 20th international conference on
  computational linguistics and intelligent text processing}, pages 7--13.

\bibitem[{Xu et~al.(2015)Xu, Ba, Kiros, Cho, Courville, Salakhudinov, Zemel,
  and Bengio}]{pmlr-v37-xuc15}
Kelvin Xu, Jimmy Ba, Ryan Kiros, Kyunghyun Cho, Aaron Courville, Ruslan
  Salakhudinov, Rich Zemel, and Yoshua Bengio. 2015.
\newblock \href {https://proceedings.mlr.press/v37/xuc15.html} {Show, attend
  and tell: Neural image caption generation with visual attention}.
\newblock In \emph{Proceedings of the 32nd International Conference on Machine
  Learning}, volume~37 of \emph{Proceedings of Machine Learning Research},
  pages 2048--2057, Lille, France. PMLR.

\bibitem[{Zhang et~al.(2022)Zhang, Chen, Wang, Deng, Zhang, Li, Wang, and
  Cao}]{zhang-etal-2022-mderank}
Linhan Zhang, Qian Chen, Wen Wang, Chong Deng, ShiLiang Zhang, Bing Li, Wei
  Wang, and Xin Cao. 2022.
\newblock \href {https://doi.org/10.18653/v1/2022.findings-acl.34} {{MDER}ank:
  A masked document embedding rank approach for unsupervised keyphrase
  extraction}.
\newblock In \emph{Findings of the Association for Computational Linguistics:
  ACL 2022}, pages 396--409, Dublin, Ireland. Association for Computational
  Linguistics.

\end{thebibliography}
\bibliographystyle{acl_natbib}

\clearpage
\appendix

\section{Appendix}
\label{sec:appendix}

\begin{table*}[b]
\centering
\arrayrulecolor{black}
\scalebox{0.85}{
\begin{tabular}{cllccc} 
\toprule
\multirow{2}{*}{Number} & \multirow{2}{*}{Encoder} & \multirow{2}{*}{Decoder}                                                                                             & \multicolumn{3}{c}{$F1@K$}  \\
\cline{4-6}                        &                          &                                                                                                                      & 5     & 10    & 15        \\ 
\midrule
1                       & Book:"[D]"               & This book mainly talks about [C]                                                                                     & \textbf{21.69} & \textbf{26.70} & \textbf{27.44}     \\
2                      & Passage:"[D]"            & This passage mainly talks about [C]                                                                                  & 21.27 & 26.15 & 27.25     \\
3                      & News:"[D]"               & This news mainly talks about [C]                                                                                     & 20.94 & 26.09 & 27.07     \\
4                      & Text:"[D]"               & This text mainly talks about [C]                                                                                     & 19.88 & 25.26 & 26.43     \\
5                      & Paper:"[D]"               & This paper mainly talks about [C]                                                                                     & 21.37 & 26.43 & 27.33     \\
\bottomrule
\end{tabular}}
\caption{Templates we design to study the impact of the noun word representing the document.}
\label{tb: noun}
\arrayrulecolor{black}
\end{table*}

\begin{table*}[b]
\centering
\arrayrulecolor{black}
\scalebox{0.88}{
\begin{tabular}{clp{10cm}ccc} 
\toprule
\multirow{2}{*}{Length} & \multirow{2}{*}{Encoder} & \multirow{2}{*}{Decoder}                                                                                             & \multicolumn{3}{c}{$F1@K$}  \\
\cline{4-6}                        &                          &                                                                                                                      & 5     & 10    & 15        \\ 
\midrule
0                       & Book:"[D]"               & {[}C]                                                                                                                & 14.40 & 14.41 & 14.99     \\
2                       & Book:"[D]"               & Book about [C]                                                                                                       & 15.38 & 20.88 & 22.84     \\
2                       & Book:"[D]"               & It is [C]                                                                                                            & 17.48 & 23.13 & 24.87     \\
2                       & Book:"[D]"               & Keywords are [C]                                                                                                     & 17.48 & 23.26 & 24.97     \\
2                       & Book:"[D]"               & Talk about [C]                                                                                                       & 15.38 & 20.88 & 22.84     \\
\midrule
5                       & Book:"[D]"               & This book are mainly about [C]                                                                                       & 21.23 & 26.28 & 27.00     \\
5                       & Book:"[D]"               & This book mainly focuses on [C]                                                                                      & 21.40 & 26.35 & 27.06     \\
5                       & Book:"[D]"               & This book mainly talks about [C]                                                                                     & 21.69 & \textbf{26.70} & \textbf{27.44}     \\
5                       & Book:"[D]"               & This book pays attention to [C]                                                                                      & 19.33 & 24.39 & 25.95     \\
\midrule
10                      & Book:"[D]"               & All in all, the core of this book is [C]                                                                             & 20.21 & 25.18 & 26.27     \\
10                      & Book:"[D]"               & Read this book and tell me that it is about [C]                                                                      & 20.25 & 25.00 & 26.46     \\
10                      & Book:"[D]"               & Take a look at the full book, it involves [C]                                                                        & 19.82 & 25.00 & 26.31     \\
10                      & Book:"[D]"               & Think carefully, this book has somthing to do with [C]                                                               & 21.27 & 26.16 & 26.93     \\
\midrule
20                      & Book:"[D]"               & Please read this book carefully from beginning to end and just give your conclusion, this book mainly focuses on [C] & 21.11 & 25.05 & 25.38     \\
20                      & Book:"[D]"               & The book describes something so interesting, please read it carefully and tell us that this book is about [C]        & 19.99 & 24.47 & 25.36     \\
20                      & Book:"[D]"               & The book is interesting, please read it carefully and summarize its main points with a few keywords like [C]         & 15.84 & 20.27 & 21.23     \\
20                      & Book:"[D]"               & Through careful reading and adequate analysis, we have come to the conclusion that this book mainly talks about [C]  & \textbf{21.89} & 26.44 & 27.11     \\

\bottomrule
\end{tabular}}
\caption{Templates we design to study the impact of template length.}
\label{tb: templates}
\arrayrulecolor{black}
\end{table*}

\subsection{Effects of the Noun Word}

We also design experiments to study the impact of the noun word representing the document (no position information used). We consistently use the best-performing template, and only vary the noun word. A total of five different words were tested. As illustrated in Table \ref{tb: noun}, the choice of noun word does affect the performance of the template, with "Book" achieving the highest results.

\subsection{Templates for the Length Study}
\label{app: length}
\noindent We use five groups of templates of different lengths to explore the effect of template length. All the templates are shown in Table \ref{tb: templates} and $F1$ here is the average of six datasets.

\end{document}